# Multimode optical fiber radiation modal decomposition

*M. V. Bolshakov[1,2], M. A. Komarova[2], N. D. Kundikova[1,2]*

[1] *Institute of Electrophysics of Urals Branch of RAS, Ekaterinburg, Russia*
[2] *South Urals State University, Chelyabinsk, Russia*
*Corresponding author: kundikovand@susu.ac.ru*

We proposed the method of the optical fiber modal decomposition of the radiation propagating in a multimode optical fiber with a step like refractive index profile. The field distribution at the output end of the fiber was used. The method is based on the field decomposition by non-orthogonal modes. The full complex expansion coefficients of the light field were calculated for theoretical data.

In comparison with a few mode fibers the multimode fibers have many advantages in the optical communication system. Multimode fiber can transmit a lot of information and parallel data transfer is possible due to the multiple modes, which present in multimode fiber. Examples are dispersive multiplexing [1], spatial mode multiplexing [2], and the transmission of spatial image through the fiber [3]. The decomposition of radiation, which propagates in the optical fiber over the modes, is an important issue in the development and researching of multimode communication systems. Nevertheless, up to date a method for universal decomposition of the field in terms of the components eigenmodes has not been realized [4-7].

In this paper we demonstrate the theoretical method of multimode optical fiber radiation mode decomposition. Method is based on the complex light field decomposition by non-orthogonal modes. Let us consider light propagation in an optical fiber with a step-index profile. In the approach of weakly directing waveguide the requirements of symmetry allow to write down four polarization modes in the following form for fiber length $z$, any orbital angular momentum $m$ ($m \geq 0$) and radial quantum number $N$ [8, 9]:

$$\begin{aligned}
\mathbf{M}_{m,N}^{(1)}(x,y,z) &= [\cos(m\varphi)\mathbf{e_x} - \sin(m\varphi)\mathbf{e_y}] \cdot F_{m,N}(x,y) \cdot e^{iz\beta_{m,N}^1}, \\
\mathbf{M}_{m,N}^{(2)}(x,y,z) &= [\cos(m\varphi)\mathbf{e_x} + \sin(m\varphi)\mathbf{e_y}] \cdot F_{m,N}(x,y) \cdot e^{iz\beta_{m,N}^2}, \\
\mathbf{M}_{m,N}^{(3)}(x,y,z) &= [\sin(m\varphi)\mathbf{e_x} + \cos(m\varphi)\mathbf{e_y}] \cdot F_{m,N}(x,y) \cdot e^{iz\beta_{m,N}^3}, \\
\mathbf{M}_{m,N}^{(4)}(x,y,z) &= [\sin(m\varphi)\mathbf{e_x} - \cos(m\varphi)\mathbf{e_y}] \cdot F_{m,N}(x,y) \cdot e^{iz\beta_{m,N}^4}.
\end{aligned} \quad (1)$$

Here $\mathbf{e_x}, \mathbf{e_y}$ are the eigenvectors, $\varphi = \arctan(x/y)$. The radial distribution functions $F_{m,N}(x,y)$ are Bessel and McDonald functions, $\beta_{m,N}^k$ are propagation constants, which determine the velocity of mode propagation.

Decomposition of an arbitrary function in the system of non-orthogonal functions is a classical problem of functional analysis. The general form of electric component decomposition of the light field $\mathbf{E}(x,y)$, propagated in the optical fiber, can be represented as a superposition of the polarization modes $\mathbf{M}_{m,N}^k(x,y)$:

$$\mathbf{E}(x,y) = \sum_{k=1}^{4} \sum_{m} \sum_{N} C_{m,N}^k \mathbf{M}_{m,N}^k(x,y), \quad (2)$$

where $C_{m,N}^k$ are complex polarization mode expansion coefficients, which determine the contribution of each modes in the total light field. The problem of multimode optical fiber radiation modal decomposition is reduced to finding expansion coefficients $C_{m,N}^k$. Let us form the system of linear equations. For convenience we introduce continuous numbering $i, j = 0 \dots (L-1)$ instead of $k, m, N$ indexes, where $(L-1)$ is number of modes, propagating in a multimode optical fiber. So Eq. 2 has a form:

$$\mathbf{E}(x,y) = \sum_i C_i \mathbf{M_i}(x,y). \quad (3)$$

The constants $C_i$ are the expansion coefficients of the light field $\mathbf{E}(x,y)$ in the basic functions $\{\mathbf{M_i}(x,y)\}$. The field distribution of each mode $\mathbf{M_i}(x,y)$ is determined from Eq. (1). The total light field is known from the experiment. Thus, the expansion coefficients $C_i$ may be easily represented in a non-orthogonal system of functions by the linear equations system

$$\sum_{i=0}^{L-1} C_i (\mathbf{M_i}, \mathbf{M_j}) = (\mathbf{E}, \mathbf{M_j}), \quad j = 0, \dots, (L-1). \quad (4)$$

Solving this linear equations system, we can obtain the expansion coefficients $C_i$. Software for calculation of complex mode expansion coefficients $C_i = a_i \cdot \exp(i\phi_i)$ has been written in Matlab. To check the method of the multimode optical fiber radiation modal decomposition the calculation series was carried out for numerically given field distribution at the output end of the fiber. The mode decomposition of radiation at the fiber input was randomly generated and the field distribution at the fiber output $E^r(x,y)$ was calculated in accordance with given expansion coefficients $C_i^r$. Then this field distribution $E^r(x,y)$ was decomposed into non-orthogonal modes $M_i(x,y)$, presented in Eq. (1). Complex mode coefficients $C_i$ were calculated by solving the linear equations system (Eq. 4). To estimate the accuracy of the used method we compared calculated expansion coefficients $C_i$ with initially given expansion coefficients $C_i^r$.

We have carried out the calculations using an optical fiber with a step like index profile and the following parameters, fiber core diameter $2\rho = 12$ $\mu$m, difference between the refraction index of the core and cladding $\delta n = n_{co} - n_{cl} = 0.004$ on the fiber length $z$ at the wavelength $\lambda = 0.6328$ $\mu$m. There are seven polarization modes $\mathbf{M_i}$ propagating in this optical fiber for given wavelength. The amplitudes $a_i$ and phases $\phi_i$ expansion coefficients of modes $\mathbf{M_i}$ are presented in the table 1. This method of fiber modal decomposition is suitable for fibers with arbitrary parameters and any amount of modes.

Table 1

The amplitudes $a_i^\pm$ and phases $\phi_i^\pm$ mode expansion coefficients for each $m, N$
(given and calculated expansion coefficients are presented the right and left, respectively)

| $m$ | $N$ | $a_i^+$ | | $\phi_i^+$ | | $a_i^-$ | | $\phi_i^-$ | |
|---|---|---|---|---|---|---|---|---|---|
| 0 | 1 | 0,45 | 0,58 | 2,53 | 1,80 | 0,45 | 0,62 | 1,92 | 2,60 |
| 0 | 2 | 0,57 | 0,47 | -2,18 | -1,60 | 0,57 | 0,81 | 1,82 | 1,50 |
| 1 | 1 | 0,11 | 0,11 | -2,20 | -2,20 | 0,25 | 0,25 | 2,80 | 2,80 |
| 1 | 2 | 0,60 | 0,60 | 2,50 | 2,50 | 0,70 | 0,70 | 1,60 | 1,60 |
| 2 | 1 | 0,24 | 0,24 | -1,50 | -1,50 | 0,50 | 0,50 | 1,80 | 1,80 |
| 3 | 1 | 0,29 | 0,29 | 2,50 | 2,50 | 0,44 | 0,44 | -1,60 | -1,60 |
| 4 | 1 | 0,65 | 0,65 | -2,80 | -2,80 | 0,75 | 0,75 | 2,10 | 2,10 |

As a result, it was found, that uncertainty of calculating expansion coefficients is 0.001% for $m > 0$, however expansion coefficients $C_i$ were different from given expansion coefficients $C_i^r$ for $m = 0$. These results are explained by the fact, that the field distribution is axisymmetric for $m = 0$ and several sets of modal expansion coefficients, which meet the condition, were found. So the modal expansion coefficients for $m = 0$ are degenerated. The obtained field distribution with calculating expansion coefficients $C_i$ was completely identical to the field distribution with given expansion coefficients $C_i^r$. The mean-square deviation of field distribution was $10^{-12}$. Numerical calculations were carried out for different numbers of modes propagating in the optical fiber (the maximum number of modes are 30) and the high accuracy of finding the mode expansion coefficients was shown. So the problem of multimode optical fiber radiation modal decomposition has been solved for the numerical field distribution at the end of a fiber. The uncertainty of the method was 0.001%.

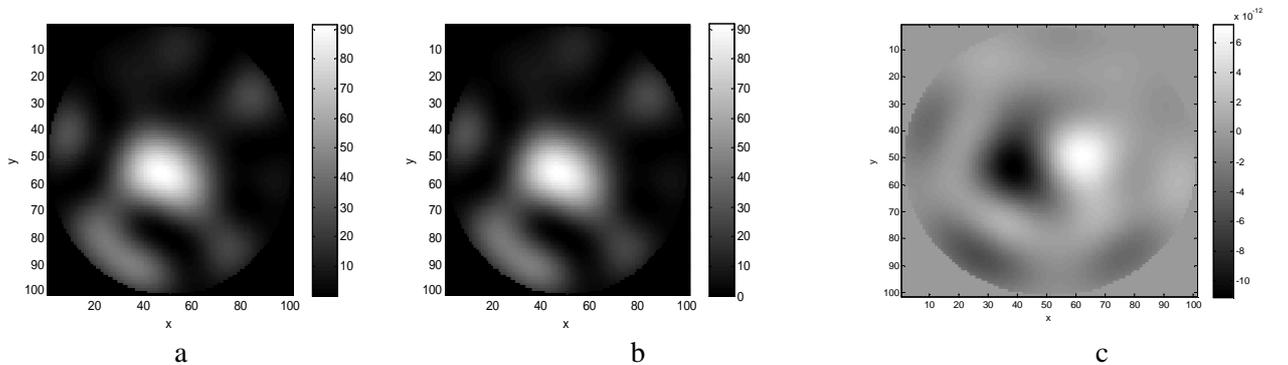

a  b  c

Fig. 1. The intensity distribution with given expansion coefficients (a), calculating expansion coefficients (b) and the distinction of this intensity distribution (c).

In conclusion, we have presented an approach to perform modal decomposition for field in optical fiber supporting any amount of modes. The obtained reconstruction results are in very good agreement with the numerically given field distribution at the output end of the fiber. This confirms the possibility of accurate measurements of the modal content. The method was proven to work well also in cases with a large amount of mode content. In the near future we plan to conduct a study with experimental data.

**Referents**